\documentclass[twocolumn]{aastex701}

\newcommand{\taud}{\ensuremath{\tau_{\rm d}}} 
\newcommand{\tauuv}{\ensuremath{\tau_{\rm 2500}}} 

\newcommand{\sigmad}{\ensuremath{\sigma_{\rm d}}} 
\newcommand{\sigmauv}{\ensuremath{\sigma_{\rm 2500}}} 

\newcommand{\lv}{\ensuremath{L_{\rm 5100}}} 
\newcommand{\z}{\ensuremath{z}} 
\newcommand{\opz}{\ensuremath{1+z}} 
\newcommand{\lamrest}{\ensuremath{\lambda_{\rm rest}}} 
\newcommand{\rfe}{\ensuremath{R_{\rm Fe}}} 
\newcommand{\mbh}{\ensuremath{M_{\rm BH}}} 
\newcommand{\mdot}{\ensuremath{\dot{M}}} 
\newcommand{\mdotdim}{\ensuremath{\dot{\mathcal{M}}}} 
\newcommand{\ruv}{\ensuremath{R_{\rm 2500}}} 
\newcommand{\msun}{\ensuremath{M_{\odot}}} 

\newcommand{\hb}{\ensuremath{\rm{H}\beta}} 
\newcommand{\mgii}{\ion{Mg}{2}} 
\newcommand{\feii}{\ion{Fe}{2}} 
\newcommand{\rs}{\ensuremath{r_{\rm s}}} 
\newcommand{\p}{\ensuremath{p}} 

\usepackage{xcolor}

\usepackage{xeCJK}        

\begin{document}


\title{Probing the Optical/UV Continuum Variability of Active Galactic Nuclei with the Damped Random Walk Model}

\correspondingauthor{Wei-Hao Bian}
\email[show]{whbian@njnu.edu.cn}

\author[0009-0004-4431-4535]{Xiang-Wei Ning(宁翔炜)}
\affiliation{School of Physics and Technology, Nanjing Normal University, Nanjing 210023, People's Republic of China}
\email{xwning@njnu.edu.cn}

\author{Wen-Jie Hu(胡文杰)}
\affiliation{School of Physics and Technology, Nanjing Normal University, Nanjing 210023, People's Republic of China}
\email{3492445582@qq.com}

\author{Zi-Han Yuan(袁子涵)}
\affiliation{School of Physics and Technology, Nanjing Normal University, Nanjing 210023, People's Republic of China}
\email{3492445582@qq.com}

\author{Yi Tang(唐毅)}
\affiliation{School of Physics and Technology, Nanjing Normal University, Nanjing 210023, People's Republic of China}
\email{1834742996@qq.com}

\author[0009-0008-5295-3977]{Yan-Song Ma(马岩松)}
\affiliation{School of Physics and Technology, Nanjing Normal University, Nanjing 210023, People's Republic of China}
\email{2244180800@qq.com}

\author{Yu-Meng Guan(官雨蒙)}
\affiliation{School of Physics and Technology, Nanjing Normal University, Nanjing 210023, People's Republic of China}
\email{2304174572@qq.com}

\author{Le-Yi Huang(黄乐奕)}
\affiliation{School of Physics and Technology, Nanjing Normal University, Nanjing 210023, People's Republic of China}
\email{1213023754@qq.com}

\author[0000-0002-2121-8960]{Wei-Hao Bian(卞维豪)}
\affiliation{School of Physics and Technology, Nanjing Normal University, Nanjing 210023, People's Republic of China}
\email{whbian@njnu.edu.cn}

\begin{abstract}
\noindent Optical/UV continuum variability over baselines of up to 20 yr is investigated using the damped random walk (DRW) model for 1214 active galactic nuclei (AGNs) compiled from Y. Shen et al.  and C. J. Burke et al. Among these, 115 AGNs with reliable DRW fitting and unbiased DRW parameters  (timescale \tauuv\ and amplitude \sigmauv\ at the rest-frame 2500~\AA) are used to examine the relations between the variability and the accretion properties. We found a significant correlation of $\tau_{2500}$ with black hole mass (\mbh) and dimensionless accretion rate (\mdotdim) of the form $\tauuv \propto \mbh^{0.50\pm0.07}\mdotdim^{0.27\pm0.04}$, while the dependence on the continuum luminosity at 5100~\AA\ and redshift becomes insignificant after controlling for \mbh. \tauuv\ also correlate with the mass accretion rate (\mdot) as $\tauuv \propto \mdot^{0.41\pm0.04}$, which is consistent with the thermal timescale expected from the standard thin-disk model within $3\sigma$ uncertainties; while the average value is consistent with the expectations of the model, the scatter in the relation can be due to differences in the viscosity parameter ($\alpha$) or in disk temperature profiles. Assuming that \tauuv\ corresponds to the accretion disk thermal timescale, the disk-emitting radius at rest-frame 2500~\AA\ (\ruv) is found to correlate with \mbh\ as $\ruv \propto \mbh^{0.53\pm0.05}$. Using the microlensing-based \ruv--\mbh\ relation of C. W. Morgan et al., $\alpha$ is statistically estimated to be $0.04\pm0.01$.

\end{abstract}



\section{Introduction} \label{sec:1}

In the current understanding of galactic structure, it is widely accepted that supermassive black holes (SMBHs) reside at the centers of most massive galaxies \citep[e.g.,][]{Kormendy_2013}. In active galactic nuclei (AGNs), the intense energy emitted across multiple wavelengths is believed to be powered by the accretion of material onto the SMBH \citep[e.g.,][]{Salpeter_1964}. Accreted material forms an accretion disk surrounding the SMBH, with the optical/ultraviolet (UV) continuum emission thought to originate primarily from the accretion disk. For a geometrically thin, optically thick disk, it is described by the standard thin-disk model \citep{Shakura_1973}. It is worth noting that the standard thin-disk model faces several challenges in explaining observations, while magnetohydrodynamics simulations can provide valuable insights into the physics of accretion disks \citep[e.g.,][]{Davis_2020}. Observations reveal that AGN emission is intrinsically variable across the electromagnetic spectrum, with variability observed on a wide range of timescales and amplitudes \citep[e.g.,][]{Hawkins_1993, Macleod_2010, Netzer_2013, Lin_2025, Petrecca_2026}. However, the physical mechanisms driving this variability are complex and debated \citep[e.g.,][]{Hawkins_1993,Hawkins_2007, Kawaguchi_1998, Peterson_2004, Li_2008, Kelly_2009, Macleod_2010, Rakshit_2017, Sanchez_Saez_2018, Lu_2019, Burke_2021, Tang_2023, Kara_2025, Paolillo_2025, Patel_2025, Xu_2025,  Petrecca_2026, Rachana_2026}.

There is a growing consensus that this variability is fundamentally stochastic in nature. The lowest-order continuous-time autoregressive moving-average (CARMA) model, i.e., the damped random walk (DRW) model, can be used as an approximation to describe AGNs optical/UV variability \citep[e.g.,][]{Kelly_2009, Macleod_2010, Zu_2013, Burke_2021, Tarrant_2025}, although some studies, owing to the limitations of the DRW model, have used second-order CARMA process or empirical models to describe AGNs optical/UV variability \citep[e.g.,][]{Yu_2022, Yu_2025, Petrecca_2026}. The two parameters of the DRW model are the damping timescale \taud, and the long-term amplitude \sigmad, where \taud\ represents the characteristic timescale over which the light curve becomes approximately uncorrelated, and \sigmad\ quantifies its long-term standard deviation \citep[e.g.,][]{Macleod_2010, Hu_2024}. 

\cite{Kelly_2009} compiled optical light curves for 100 AGN/quasars from the literature. Among these, 20 AGNs have virial black hole mass (\mbh) measured via reverberation mapping (RM), and 51 AGNs have single-epoch (SE) virial mass estimates. With DRW, they found that the $R$-band damping timescale \taud\ correlates with the continuum luminosity at 5100 \AA\ (\lv), \mbh, and the Eddington ratio ($\lambda_{\rm Edd} \equiv L_{\rm bol}/L_{\rm Edd}$, where $L_{\rm bol}$ and $L_{\rm Edd}$ are bolometric luminosity and Eddington luminosity), following $\taud \propto \lv^{0.29\pm0.08}$, $\taud \propto \mbh^{0.56\pm0.14}$ and $\taud \propto \lambda_{\rm Edd}^{-0.06\pm0.27}$. They also found that the amplitude of short-timescale variability is significantly anticorrelated with \mbh\ and \lv, and no evidence for a dependence of variability amplitude on $\lambda_{\rm Edd}$.
\cite{Macleod_2010} analyzed 10 yr of photometric data for $\sim 9000$ quasars from the Sloan Digital Sky Survey (SDSS) Stripe 82 to investigate the correlations between DRW parameters and AGN properties, including rest-frame wavelength (\lamrest), \mbh, $i$-band magnitude ($M_i$), and $\lambda_{\rm Edd}$. They found $\taud \propto \lamrest^{0.17\pm0.02} M_i^{0.03\pm0.04} \mbh^{0.21\pm0.07}$, $\mathrm{SF}_\infty \propto \lamrest^{-0.479\pm0.005} M_i^{0.131\pm0.008} \mbh^{0.18\pm0.03}$ and $\rm SF_\infty \propto \lambda_{Edd}^{-0.23\pm0.03}$, where $\rm SF_\infty = \sqrt{2}\sigmad$ \citep[also see][for 15 yr baseline light curves]{Suberlak_2021}. \cite{Lu_2019} compiled 73 RM AGNs with spectroscopic 5100~\AA\ light curves to investigate the correlations of DRW parameters with \mbh, \lv, and the dimensionless accretion rate (\mdotdim). They found $\taud \propto \lv^{0.46\pm0.09}$, $\taud \propto \mbh^{0.60\pm0.09}$, and $\Sigma_{\rm d} \propto \mbh^{0.20\pm0.04}$, $\Sigma_{\rm d} \propto \mdotdim^{-0.13\pm0.03}$, where $\Sigma_{\rm d}$ is \sigmad\ normalized by the mean flux. \cite{Burke_2021} compiled 67 broad-line AGNs with \mbh\ in the range $\sim 10^4-10^{10} M_\odot$ (from RM or SE virial methods) and found $\taud \propto \mbh^{0.38\pm0.05}$. They also found a relation between the accretion disk size at 2500~\AA\ and \mbh. With additional 78 low-mass AGNs, \cite{Wang_2023} provided further confirmation of this \taud--\mbh\ relation and suggested a method of direct \mbh\ estimation from \taud. For a jet-related AGNs sample of 34 blazars and seven microquasars, \cite{Zhang_2024} found $\taud \propto \mbh^{0.57\pm0.02}$, implying a possible disk-jet connection based on time-domain analysis.

The power spectrum density (PSD) or the structure function is commonly used to characterize the variability properties of AGNs \citep[e.g.,][]{Tang_2023, Son_2025, Petrecca_2026}. When the DRW model is assumed, the structure function can be expressed as $\mathrm{SF}(\Delta t)=\mathrm{SF}_\infty(1-e^{-|\Delta t|/\taud})^{1/2}$, where $\Delta t$ is the time interval, and the damping timescale \taud\ corresponds to the bending timescale in the PSD or structure function \citep[e.g.,][]{Macleod_2010, Kozlowski_2016}. Using the structure function of $\sim 5000$ quasars, \cite{Tang_2023} found a universal structure function when the time interval is expressed in units of the thermal timescale from the standard thin-disk model and implied the magnetorotational instabilities as a main cause of the variability. Using Zwicky Transient Facility (ZTF) $g$-band light curves for 4770 SDSS quasars ($0.6<z<0.7$), \cite{Arevalo_2024} found PSD breaking timescale $t_b \propto \mbh^{0.65-0.55} \lambda_{\rm Edd}^{0.35-0.3} $, and suggested a universal PSD. For quasars from the SDSS-RM project, \cite{Son_2025} employed the ensemble structure function and found that the characteristic timescale scales as $L_{\rm bol}^{0.50\pm0.03}\lamrest^{1.42\pm0.09}$, assuring a single universal structure function. For the light curves ($\sim$ 20 yr) in the $r$ band for the 4037 quasars, \cite{Petrecca_2026} used a bending power law (BPL) to model the ensemble PSD. They found that the BPL amplitude has an anticorrelation with $\lambda_{\rm Edd}$ and does not depend on \mbh. They also found that the bending frequency ($\sim 1/\taud$) has a relation with \mbh\ and not $\lambda_{\rm Edd}$. The relation of variability with AGN properties or the difference in slope $k$ for $\taud \propto \mbh^{k}$ from different samples is still a question open to debate \citep[e.g.,][]{Paolillo_2025}.

There remains an ongoing debate regarding how to obtain an unbiased estimate of DRW $\taud$. Several studies proposed criteria based on simulations \citep[e.g., ][]{Kozlowski_2017, Hu_2024}. \cite{Kozlowski_2017} suggested $D/\taud > 10$ \citep[see also][]{Lu_2019, Burke_2021, Wang_2023, Chen_2025, Xiong_2025}, where $D$ is the light-curve baseline, whereas \cite{Kozlowski_2021} suggested $D/\taud > 30$ is necessary. In contrast, \cite{Suberlak_2021} claim that $D/\taud \geq 3$ is sufficient to recover \taud\ without substantial bias \citep[see also][]{ Stone_2022}. \cite{Hu_2024} found that the criteria for unbiased \taud\ rely on statistical assumptions.

Several theoretical models have been proposed to explain the optical/UV continuum variability observed in AGNs. A widely considered model involves accretion disk instability; it posits that episodic flare events within the disk can result in significant changes in luminosity \citep[e.g.,][]{Hawkins_1993, Kawaguchi_1998}. Changes in accretion rate and microlensing effects can also drive optical/UV variability in AGNs \citep[e.g.,][]{Pereyra_2006, Li_2008, Blackburne_2010, Zheng_2024}. Another explanation is the X-ray reprocessing model, in which variable X-ray emission from the corona illuminates the accretion disk and induces variations in the optical/UV emission \citep[e.g.,][]{Cackett_2021, Papoutsis_2024}. \cite{Sun_2020} proposed the Corona-heated Accretion-disk Reprocessing (CHAR) model to explain the optical/UV variability of AGN. In this model, the accretion disk and corona are tightly coupled through magnetic fields. Magnetohydrodynamic fluctuations in the corona influence disk turbulence and alter the local heating rate, leading to temperature fluctuations in the disk that ultimately drive the observed optical/UV flux variations. \cite{Zhou_2024} simulated light curves according to the CHAR model and found that \taud\ is strongly correlated with the bolometric luminosity and the rest-frame wavelength ($\taud \propto L_{\rm bol}^{0.65}\lamrest^{1.19}$), but shows little or no correlation with \mbh.

In comparing the observational results and theoretical predictions from SMBH accretion-disk model on AGN variability, \mbh\ is a key parameter, as is the accretion rate. For broad-line AGNs, the broad-line region (BLR) clouds can be used as a probe to calculate the virial mass of the black hole \citep[e.g.,][]{Peterson_2004, Netzer_2013}:
 \begin{equation}
 \label{eq1}
\mbh=f\times \frac{R_{\rm BLR}~(\Delta V)^2}{G}.
\end{equation}
where $\Delta V$ is the velocity of the BLR clouds traced by the the full width at half-maximum ($\rm FWHM$) or the line dispersion for the broad \hb\ emission line, $f$ is a corresponding virial factor \citep[e.g.,][]{Peterson_2004, Ho_2014, Yu_2020mass, Li_2025}, and $G$ is the gravitational constant. The BLRs size, $R_{\rm BLR}$, can be measured by the AGNs RM method or the empirical $R_{\rm BLR} - \lv$ relation for the broad \hb\ emission line \citep[e.g.,][] {Bentz_2013, Du_2019, Yu_2020}. Other broad lines (H$\alpha$, \ion{Mg}{2}, \ion{C}{4}) are also used where the broad \hb\ is unavailable \citep[e.g.,][]{Shen_2011}. However, it is suggested that there are systematic biases in the \ion{C}{4}-based \mbh\ \citep[e.g.,][]{Shen_2011, Bian_2012, Shen_2013, Shen_2024}. 

In this paper, we use 115 light curves with up to a 20 yr baseline with reliable DRW model fitting and unbiased DRW parameters to investigate the relations between the optical/UV continuum variability (DRW timescale and amplitude) and the SMBH accretion properties. This paper is organized as follows. Section~\ref{sec:2} describes the sample and introduces the DRW model. Section~\ref{sec:3} presents the results, and Section~\ref{sec:4} discusses their implications. The main conclusions are summarized in Section~\ref{sec:5}.


\section{Sample and methods} \label{sec:2}

\subsection{Long-term AGN Light Curves}
Our investigation of long-term continuum variability in AGNs begins with a sample that includes light curves from 1214 quasars/AGNs, compiled from the SDSS-RM project \citep{Shen_2024} and \cite{Burke_2021}. For the SDSS-RM project, optical spectroscopic monitoring was obtained with the BOSS multiobject spectrographs on the SDSS 2.5 m wide-field telescope at Apache Point Observatory. The photometry in the $g$ and $i$ bands was obtained using the Steward Observatory Bok 2.3 m telescope on Kitt Peak and the 3.6 m Canada-France-Hawaii Telescope (CFHT) on Maunakea. In addition to dedicated Bok and CFHT observations, photometry from the 1.8 m Pan-STARRS1 (PS1) telescope in the $g$, $r$, and $i$ bands was incorporated. Photometric light curves from the ZTF on the 48-inch (1.2 m) telescope were also included when available. The PyCALI code \citep{Li_2014}, which is based on the DRW model, is used to intercalibrate light curves from different facilities and bands to the synthetic SDSS $r$-band flux derived from SDSS spectroscopy. Interband continuum lags are ignored, as they are typically smaller than the BLR lags \citep{Shen_2024}. The synthetic broadband flux computed from the SDSS spectroscopy in the $r$-band serves as the baseline to calibrate and merge other photometric light-curve sets.
We use merged light curves scaled to the r-band flux from SDSS spectra provided by the SDSS-RM project \footnote{\url{https://ariel.astro.illinois.edu/sdssrm/final_result/}} for 849 broad-line quasars over a redshift range of $0.1 < z < 4.5$ \citep{Shen_2024}. These light curves span an 11 yr baseline from 2010 to 2020, with a mean cadence of $\sim 5$ days in the observed frame. In addition to the SDSS-RM light curves, we also used 414 optical light curves for 365 AGNs\footnote{\url{https://github.com/burke86/taufit/tree/master/paper}} in different bands (e.g., $g$ and $r$ bands) compiled by \cite{Burke_2021}. These light curves are from serval surveys, including: (1) multiple facilities for the SDSS Strip 82 quasars (e.g., SDSS, PS1, ZTF $g$-band), (2) various RM programs (e.g., 5100~\AA\ light curve of the International AGNs Watch and RM of super-Eddington accreting massive black holes), (3) wide-area time-domain imaging surveys (e.g., ZTF $r$-band). Most AGNs (190 out of 365) are quasars from SDSS Strip 82 with a 20 yr long baseline (extended using data from multiple facilities) and a mean cadence of $\sim 30$ days in the observed frame.

\subsection{DRW Model} 

\begin{figure*}
\epsscale{1.0}
\plotone{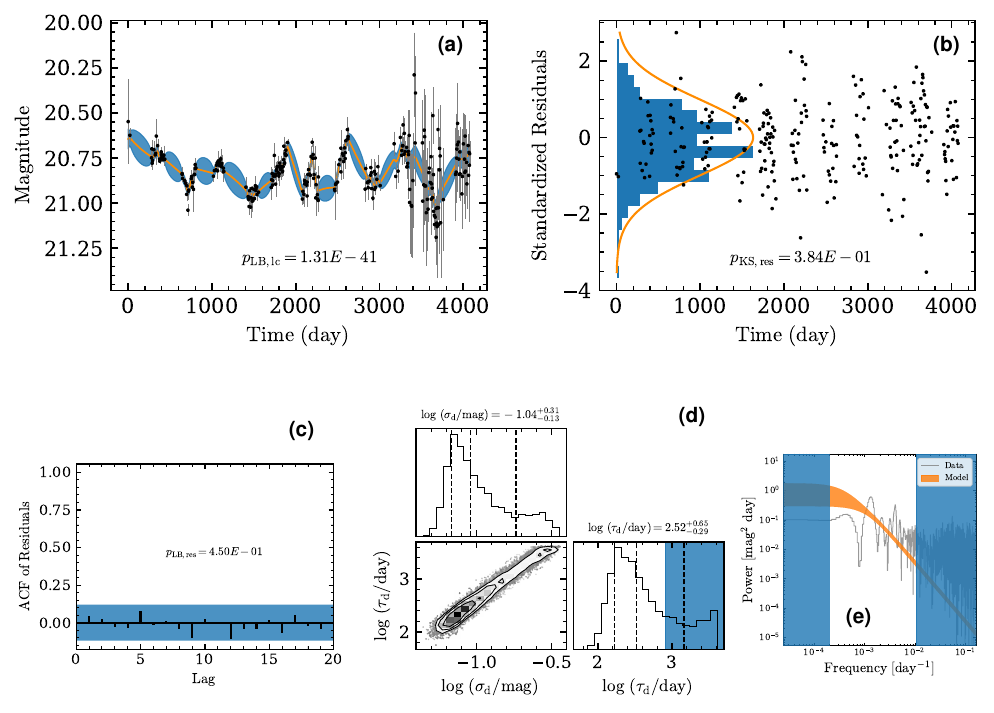}
\caption{An example of the DRW model for the long-term (about 11 yr) light curve of RM443 ($z=1.12$). Panel (a): $r$-band light curve (black points) and the DRW best fit (orange solid line) with $1\sigma$ uncertainty (blue shaded region). The $p$-value of the LB test for the light curve ($p_{\rm LB,lc}$) is shown. Panel (b): standardized residuals (black points) and their distribution (blue histogram) as well as the standard normal distribution (orange solid line). The p-value of the Kolmogorov--Smirnov test comparing the standardized residuals to a standard normal distribution ($p_{\rm KS,res}$) is shown. Panel (c): autocorrelation function of the standardized residuals, the blue shaded region represents the 95\% confidence region for white noise. The $p$-value of LB test for the standardized residual ($p_{\rm LB,res}$) is shown. Panel (d): the posterior probability distributions for the DRW parameters; the blue shaded region represents $\taud > 0.2\times D$. Panel (e): PSD of the light curve and the DRW model prediction. The gray line represents the PSD calculated from the light curve. The orange shaded region represents the $1\sigma$ uncertainty of the DRW model prediction. The blue shaded region indicates frequencies corresponding to periods longer than 20\% of the light-curve baseline and shorter than the mean cadence. Frequencies are in units of $1/(2 \pi t)$.}
\label{fig:1}
\end{figure*}

\begin{figure*}
\epsscale{1.0}
\plotone{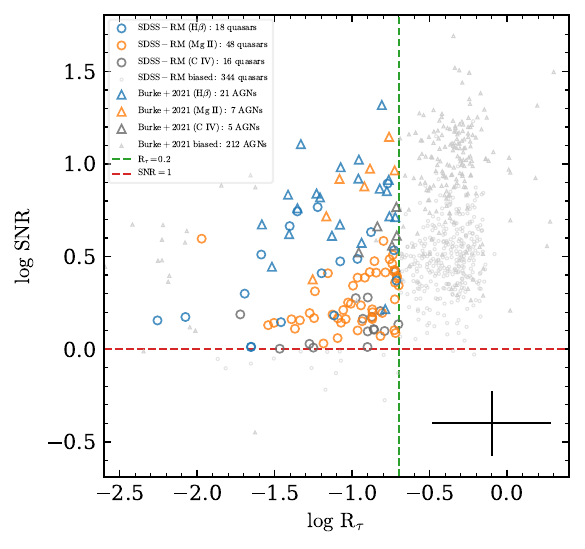}
\caption{$\rm SNR$ vs. $R_\tau$ for  671 AGNs with reliable DRW fitting, where $\rm SNR=\sigmad/\langle \delta m \rangle$ and $R_\tau=\taud/D$,  $D$ is the light-curve baseline. The green and red dashed lines represent $R_\tau=0.2$ and $\rm SNR=1$, respectively. Circles and triangles denote AGNs from \cite{Shen_2024} and \cite{Burke_2021}, respectively.
 The gray background circles and triangles represent 556 ANGs that do not satisfy the selection criteria for the unbiased DRW parameters described in Section \ref{sec:2}. Blue, orange, and gray represent \hb, \ion{Mg}{2}, and \ion{C}{4}-based virial \mbh, respectively. The number of AGNs in each category is indicated in the legend. The black cross in the lower-right corner represents $1\sigma$ mean uncertainties in the X and Y data.}
\label{fig:2}
\end{figure*}

The DRW is used to model long-term light curves for our entire sample of 1214 quasars/AGNs (849 SDSS-RM quasars from \citealp{Shen_2024} and 365 AGNs from \citealp{Burke_2021}). The DRW model describes variability as a Gaussian process characterized by a covariance (or kernel) function of the form \cite[e.g.,][]{Kelly_2009, Burke_2021, Aigrain_2023}
\begin{equation}
k(t_{n},t_{m}) = 2\sigma_{\rm d}^2\exp\left(-\frac{|t_{n}-t_{m}|}{\tau_{\rm d}}\right),
\end{equation}
where $t_{n}$ and $t_{m}$ denote different epochs, \sigmad\ is the variability amplitude, and \taud\ is the damping timescale. The DRW parameters of the light curves are obtained using \texttt{CARMA\_PACK}\footnote{\url{https://github.com/brandonckelly/carma_pack}} with $\texttt{nsamples}=60000$ \citep{Kelly_2014, Stone_2022}. The DRW model is used to fit the long-term light curves in units of magnitude as suggested by \cite{Kelly_2009}. For our sample of 1214 AGNs, the flux density ($f_\nu$) is converted to AB magnitude, $m = -2.5\log f_\nu[\rm erg~s^{-1}~cm^{-2}~Hz^{-1}] - 48.6$, with uncertainty $\delta m = \frac{2.5}{\ln(10)} \frac{\delta f_\nu}{f_\nu}$, where $\delta f_\nu$ is the flux density uncertainty  \citep[e.g.,][]{Fukugita_1996}.

We remove outliers and data points with large uncertainties from the light curves as follows. Hampel filter (implemented in \texttt{taufit}\footnote{\url{https://github.com/burke86/taufit}}) is used to remove data points that deviate from the median by more than 1.4826$\times$3 times the median absolute deviation (MAD; the factor 1.4826 scales the MAD to a Gaussian distribution) within a 300 day sliding window. We also remove data points with uncertainties that deviate from the median uncertainty by more than 3 times the standard deviation of the light-curve uncertainties \citep[e.g.,][]{Burke_2021, Xiong_2025}. This procedure is iteratively repeated until no additional data points are removed. To reduce the impact of short-term noise and improve the signal-to-noise ratio (SNR), we further apply a 5 day binning to the light curves. For each bin, the time is taken as the median, and the magnitude is calculated as the weighted mean, $m_{\rm bin}=\sum_i m_i \delta m_i^{-2}/\sum_i \delta m_i^{-2}$, with uncertainty $\delta m_{\rm bin}=\sqrt{1/\sum_i\delta m_i^{-2}}$, where $m_i$ and $\delta m_i$ are the $i$th measured magnitude and its uncertainty for each data point within the bin \citep[e.g.,][]{Bevington_2003}. An example of the DRW fit for the light curve of RM443 (from SDSS-RM) is shown in Figure~\ref{fig:1}. Panel (a) is the $r$-band light curve and the best-fit DRW model with $1\sigma$ uncertainty. Panel (b) shows the distribution of the standardized residuals \citep[details of the standardized residuals see][]{Kelly_2014}. Panel (c) is the autocorrelation function of the standardized residuals. Panel (d) shows the posterior probability distributions of the DRW parameters. The parameter values are taken as the 50th percentile of the distribution, with uncertainties given by the 16th and 84th percentiles. Panel (e) shows the PSD calculated from the light curve and the PSD predicted by the DRW model. Four additional examples with the smallest and largest $\taud$ and $\sigmad$ are provided in Figures \ref{fig:11}, \ref{fig:12}, \ref{fig:13}, and \ref{fig:14} in the Appendix \ref{sec:appendix}. The complete figure set (115 images) is available in the online journal.

To assess the goodness of fit of the DRW model, we adopting the following criteria \citep[e.g.,][]{Kelly_2009, Kelly_2014, Zhang_2024, Xiong_2025}: (1) the standardized residuals are consistent with standard normal distribution, i.e., the $p$-value of the Kolmogorov--Smirnov test\footnote{\url{https://docs.scipy.org/doc/scipy/reference/generated/scipy.stats.kstest.html}} comparing the standardized residuals to a standard normal distribution is larger than 0.05 under the null hypothesis that they are drawn from the same distribution, (2) the standardized residuals are consistent with white noise, i.e., the $p$-value of the Ljung-Box (LB) test\footnote{\url{https://www.statsmodels.org/dev/generated/statsmodels.stats.diagnostic.acorr_ljungbox.html}} for the standardized residuals is larger than 0.05 under the null hypothesis of white noise. The maximum lag used for the LB test is 20 \citep[e.g.,][]{Xiong_2025}. Based on these criteria, 671 out of 1214 AGNs (55\%) are well fitted by the DRW model. Independently, 167 and 518 AGNs fail to satisfy criteria (1) and (2), respectively. Some mismatches between the data and model PSDs may arise from several factors, such as possible deviations from the DRW model and the intrinsic stochastic nature of variability, which makes a single, limited-length light curve differ from the expected behavior \citep{Allevato_2013, Stone_2022}.

For the 671 AGNs that satisfy the goodness-of-fit criteria, we further select AGNs with unbiased DRW parameters that satisfy the following criteria \citep[e.g.,][]{Burke_2021, Stone_2022, Wang_2023, Xiong_2025}: (1) $\mathrm{R}_\tau=\taud/D < 0.2$, (2) $\rm SNR=\sigmad/\langle \delta m \rangle >1$, where $\langle \delta m \rangle$ is the mean uncertainty of the light curve, (3) $\rm \taud > cadence$, (4) the light curve is inconsistent with white noise, i.e., the $p$-value of the LB test for the light curve is less than 0.05. Under these criteria, the DRW parameters should be free from biases. The objects in the $R_\tau-\rm SNR$ plane are shown in Figure~\ref{fig:2}. Circles and triangles denote AGNs from \cite{Shen_2024} and \cite{Burke_2021}, respectively. Gray background circles and triangles represent 543 AGNs that do not satisfy the selection criteria for the unbiased DRW parameters. The remaining symbols with blue, orange, and gray colors represent \hb, \ion{Mg}{2}, and \ion{C}{4}-based virial \mbh\ respectively. A total of 543 AGNs are excluded by these criteria; the number of AGNs excluded by each criterion independently is as follows: 519 by criterion (1), 18 by criterion (2), 19 by criterion (3), and 18 by criterion (4), which underscores the importance of extending the light curves to a longer baseline in the future. Our final sample comprises 115 AGNs with reliable DRW fitting and unbiased DRW parameters.


\section{Results} \label{sec:3} 

\subsection{The SMBH Accretion-related Parameters}

For our sample of 115 AGNs, the virial \mbh\ from the broad lines are estimated (\hb, \mgii, and \ion{C}{4}) (see Equation (\ref{eq1})). 
For RM AGNs, $R_{\rm BLR}$ is from the rest-frame time lag of the emission line and the corresponding line dispersion measured from the root-mean-square (rms) spectra, and a virial factor of $4.17$ is adopted \citep{Ho_2014, Yu_2020mass, Burke_2021, Shen_2024}.
For other un-RM AGNs, an empirical $R_{\rm BLR}-\lv$ relation can be used and the virial SE \mbh\ can be calculated as \citep{Vestergaard_2006, Shen_2011, Shen_2024}
\begin{eqnarray}
 \label{eq2}
\log~(\mbh/\msun)&=a+b\log~(\lambda L_{\lambda}/10^{44} \rm erg~s^{-1}) \\ &+2\log~(\rm FWHM/km~s^{-1}) \nonumber
\end{eqnarray}
where $\lambda L_{\lambda}$ is the continuum luminosity at the wavelength of $\lambda$ near the corresponding broad line, the coefficients $a$ and $b$ are empirically calibrated against local AGNs with RM masses \citep[e.g.,][]{Ho_2014, Yu_2020mass}.
For redshifts $z<0.7$, $0.7 \leq z < 1.9$, and $z \geq 1.9$, the viral \mbh\ is estimated using \hb, \mgii, and \ion{C}{4} broad emission lines, where $\lambda$ are 5100, 3000, 1350 \AA, and ($a,\ b$) are (0.91, 0.50), (0.74, 0.62), (0.66, 0.53), respectively \citep{Shen_2011, Shen_2024}. We add a systematic uncertainty of 0.3 dex in quadrature to the \mbh. Our selected AGNs span an \mbh\ range of $\sim$ $10^6 - 10^{10} M_\odot$. 

Using the standard thin-disk model, from \lv, the mass accretion rate (\mdot) and the dimensionless accretion rate (\mdotdim) are calculated as \citep[e.g.,][]{Davis_2011, Du_2014, Wang_2014b, Liu_2022}
\begin{equation}
\mdot = 0.53 \left( \frac{l_{44}}{\mathrm{cos}~i} \right)^{3/2}M_7^{-1}~M_\odot \rm~yr^{-1},
\end{equation}

\begin{equation}
\mdotdim = \frac{\dot{M}}{L_{\rm Edd}c^{-2}} = 20.1\left(\frac{l_{44}}{\mathrm{cos}~i}\right)^{3/2}M_7^{-2}, 
\end{equation}
where $l_{44} = \lv/(10^{44} \rm~erg~s^{-1})$, $M_7 = \mbh/(10^7M_\odot)$, $i$ is the inclination angle between the line of sight and the accretion disk normal axis, $\mathrm{cos}~i=0.75$ is adopted, $L_{\rm Edd}=1.5 \times 10^{38}(\mbh/M_\odot) \rm~erg~s^{-1}$ is the Eddington luminosity, and $c$ is the speed of light. There exist strong correlations between \mdotdim\ and $\lambda_{\rm Edd}$ with $\lambda_{\rm Edd}=\eta \mdotdim$, where $\eta$ is radiative efficiency \citep{Liu_2022}. For SDSS-RM AGNs with $z < 0.7$, the host-galaxy contamination at \lv\ is corrected using the 2014 spectra \citep{Shen_2024}. For AGNs compiled by \cite{Burke_2021} with $z < 0.7$, the host-galaxy contamination at \lv\ is corrected using the empirical correction of \cite{Shen_2011} when the original studies did not account for the host contribution. For AGNs with $z \geq 0.7$, using the bolometric corrections $BC_{5100}=9.26, BC_{3000}=5.15$, and $BC_{1350}=3.81$ \citep[e.g.,][]{Shen_2011, Ma_2026}, we convert $L_{3000}$ (for $0.7 \leq z < 1.9$) and $L_{1350}$ (for $z \geq 1.9$) to $L_{5100}$ to calculate \mdot\ and \mdotdim.

It was suggested that there is a strong relation between \mdotdim\ and the optical \feii\ strength relative to the broad \hb\ (\rfe) \citep[e.g.,][]{Du_2016a, Liu_2022}. \rfe\ is defined as the ratio of the rest-frame equivalent width (EW) of the optical \feii\ within 4434–4684 \AA\ to the rest-frame EW of the broad \hb, $\rfe = \rm EW(Fe\,\textsc{ii}_{4434-4684})/\rm EW(H\beta_{\rm broad})$ \citep[e.g.,][]{Boroson_1992, Shen_2014}.
For SDSS-RM quasars, \rfe\ are calculated from data in \cite{Shen_2019}. For AGNs from \cite{Burke_2021}, \rfe\ are calculated from data in \cite{Shen_2011} and \cite{Yu_2020}. 

For 115 AGNs satisfying the selection criteria described in Section \ref{sec:2}, there are 39 (25 RM- and 14 SE-based), 55 (4 RM- and 51 SE-based), and 21 (2 RM- and 19 SE-based) AGNs with \hb, \ion{Mg}{2}, and \ion{C}{4}-based virial \mbh, respectively. It is suggested that there are systematic biases in the \ion{C}{4}-based \mbh, which is more affected by a nonvirial component, e.g., outflow or disk wind \citep[e.g.,][]{Shen_2011, Bian_2012, Shen_2013, Shen_2024}. Excluding \ion{C}{4}-based \mbh, 94 AGNs with unbiased \hb\ or \mgii-based \mbh\ are selected in the analyses that involve \mbh. Otherwise, the full sample of 115 AGNs at redshifts $0<z<3.2$ (82 from \cite{Shen_2024} and 33 from \cite{Burke_2021}) is considered. There are 14 narrow-line Seyfert 1 galaxies (NLS1s) with the broad \hb\ FWHM less than $\rm 2000~km~s^{-1}$ \citep[e.g.,][]{Paliya_2024, Rachana_2026, Wu_2026}.

\subsection{ $\tau_{\rm d,rest}$ versus $\sigma_{\rm d}$}

\begin{figure*}
\plotone{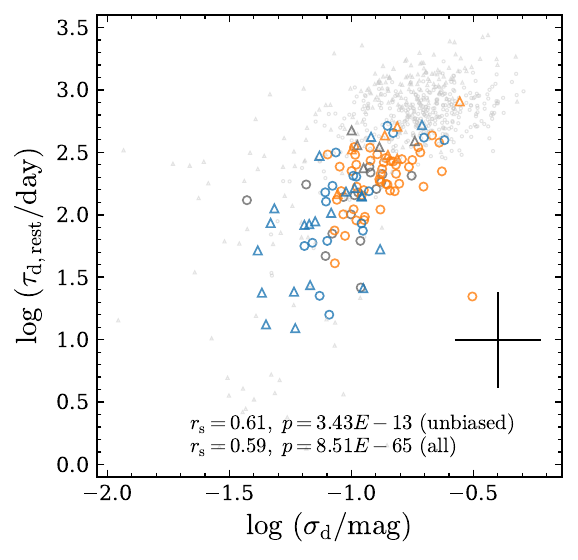}
\caption{The rest-frame damping timescale versus the variability amplitude for the 671 AGNs with reliable DRW fitting. The symbols are the same as in Figure~\ref{fig:2}. The Spearman correlation test gives $\rs=0.59$ and $\p=8.51\times10^{-65}$ for the 671 AGNs, and $\rs=0.61$ and $\p=3.43\times10^{-13}$ for our sample of 115 AGNs with unbiased DRW parameters.}
\label{fig:3}
\end{figure*}


The observed-frame damping timescale $\tau_{\rm d,obs}$ is transformed to the rest-frame at $\lamrest=\lambda_{\rm obs}/(1+z)$ as $\tau_{\rm d,rest}=\tau_{\rm d,obs}/(1+z)$. Figure~\ref{fig:3} shows the rest-frame damping timescale $\tau_{\rm d,rest}$ versus the variability amplitude \sigmad\ for the 671 AGNs with well-fitted DRW parameters. For our sample of 115 AGNs, the rest-frame damping timescale is less than about $800$ days. There is a correlation between $\tau_{\rm d,rest}$ and \sigmad. It indicates that AGNs with larger variability amplitude also have longer characteristic timescales, which is consistent with the result of \cite{Macleod_2010}. It is worth noting that the two DRW parameters can be degenerate (panel (d) of Figure~\ref{fig:1}) when the light-curve duration is not much longer than \taud\ \citep{Macleod_2010}. In such cases, the two parameters may not be independently well constrained. Thus, part of the $\sigmad-\taud$ correlation may be artificial. There are some contributions to the scatter in the variability parameters \sigmad\ and $\tau_{\rm d,rest}$, such as fitting errors, trends with physical parameters, and other sources of variability that are not captured by the model of \cite{Macleod_2010}. The Spearman correlation coefficient\footnote{\url{https://docs.scipy.org/doc/scipy/reference/generated/scipy.stats.spearmanr.html}} (\rs) and the $p$-value (\p) under the null hypothesis of no correlation between them are $\rs=0.59$ and $\p=8.51\times10^{-65}$ for all 671 AGNs, and $\rs=0.61$ and $\p=3.43\times10^{-13}$ for our sample of 115 AGNs.

\subsection{The Variability Parameters at 2500~\AA}
For a sample of $\sim 9000$ quasars in SDSS Strip 82, with light curves over 10 yr in multiple bands ($ugriz$), \cite{Macleod_2010} reported the relation between the variability timescale and the rest-frame wavelength, as well as the relation between the variability amplitude and the rest-frame wavelength. Adopting the rest-frame wavelength dependence reported by \cite{Macleod_2010}, for our sample of 115 AGNs, the DRW parameters from light curves in different single bands are converted to a common rest-frame wavelength of 2500~\AA\ following \cite{Burke_2021}. For the rest-frame timescale of $\tau_{\rm d,rest}=\tau_{\rm d,obs}/(1+z)$ at $\lamrest=\lambda_{\rm obs}/(1+z)$, the timescale is scaled to the rest-frame 2500~\AA\ as $\tauuv = \tau_{\rm d,rest}\times(\rm 2500~\AA/\lamrest)^{0.17}$ \citep{Macleod_2010}. The error of \tauuv\ is adopted to be the same as $\tau_{\rm d,rest}$.
For the $r$-band light curves of SDSS-RM quasars, $\lambda_{\rm obs}$ is adopted as the $r$-band central wavelength of 6250~\AA\ \citep{Macleod_2010}, while for the light curves of the AGNs from \citet{Burke_2021}, $\lambda_{\rm obs}$ is adopted from their Table~S1 (i.e., $\lambda_{\rm obs}$ for different filters are 8500, 4371, 5478, 6504, 4702, and 6175 \AA\ in the $TBVRgr$ bands). For the variability amplitude \sigmad, we can scale $\sigmad$ to the rest-frame 2500~\AA\ as $\sigmauv = \sigmad \times (2500\rm \AA/ \lamrest)^{-0.479}$ \citep{Macleod_2010}. The error of \sigmauv\ is adopted as the same as \sigmad. This relation is consistent with the prediction from the accretion-disk model by \cite{Li_2008}.

\subsection{The Relations between Damping Timescale and the AGN Properties} 

\begin{deluxetable*}{lcccc}
\tablecaption{
Correlations between \tauuv\ and various AGN properties.
\label{tab:1}
}
\tablehead{
\colhead{Properties} & \colhead{$\rs$} & \colhead{$\p$} & \colhead{$k$} & \colhead{$\chi_\nu^2$}
}
\startdata
$\log \lv$ & \textbf{0.43} & \boldmath{$1.98\times10^{-6}$} & $0.32\pm0.04$ & 1.69 \\
$\log \rfe$ & 0.10 & 0.49 & --- & --- \\
$\log \mbh$ & \textbf{0.45} & \boldmath{$5.12\times10^{-6}$} & $0.53\pm0.06$ & 1.38 \\
$\log \mdot$ & \textbf{0.30} & \boldmath{$3.70\times10^{-3}$} & $0.41\pm0.04$ & 1.05 \\
$\log \mdotdim$ & $-0.07$ & 0.50 & --- & --- \\
$(\log \mbh,\ \log \mdotdim)$ & \textbf{0.49} & \boldmath{$5.59\times10^{-7}$} & $(0.50\pm0.07,\ 0.27\pm0.04)$ & 0.63 \\
\enddata
\tablecomments{
Listed are the Spearman correlation coefficient (\rs), $p$-value (\p), best-fit slope ($k$), and reduced chi-square ($\chi_\nu^2$) for the correlations between $\log \tauuv$ and various AGN properties. Significant correlations with $\p < 0.01$ are highlighted in bold.
}
\end{deluxetable*}

\begin{deluxetable*}{cccc}
\setlength{\tabcolsep}{2pt}
\tablecaption{Partial correlation analyses for correlations with \tauuv. \label{tab:2}}
\tablehead{
\colhead{Parameter Pair} & \colhead{Controlled Variable} & \colhead{$\rs$} & \colhead{$\p$} 
}
\startdata
$\tauuv$ -- $\mbh$ & $\mdot$ & \textbf{0.43} & \boldmath $1.40\times10^{-5}$ \\
$\tauuv$ -- $\mdot$ & $\mbh$ & \textbf{0.2666} & \textbf{0.0098} \\
$\tauuv$ -- $\lv$ & $\mbh$ & 0.258 & 0.012 \\
$\tauuv$ -- $1+z$ & $\mbh$ & 0.22 & 0.03 \\
$\tauuv$ -- $\lv$ & $1+z$ & \textbf{0.272} &\textbf{0.003} \\
$\tauuv$ -- $1+z$ & $\lv$ & 0.01 & 0.93 \\
\enddata
\tablecomments{Listed are the Spearman correlation coefficient (\rs) and p-value (\p) of the partial correlation analyses. Significant correlations with $\p<0.01$ are highlighted in bold.}
\end{deluxetable*}

\begin{figure*}
    \centering
    \includegraphics[width=0.32\textwidth]{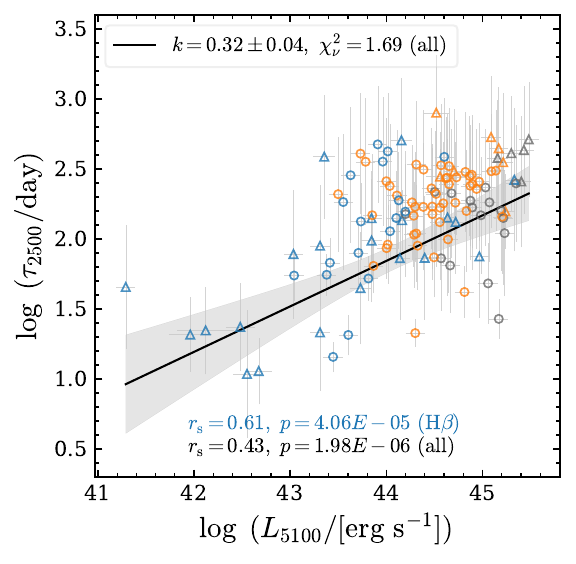}
    \includegraphics[width=0.32\textwidth]{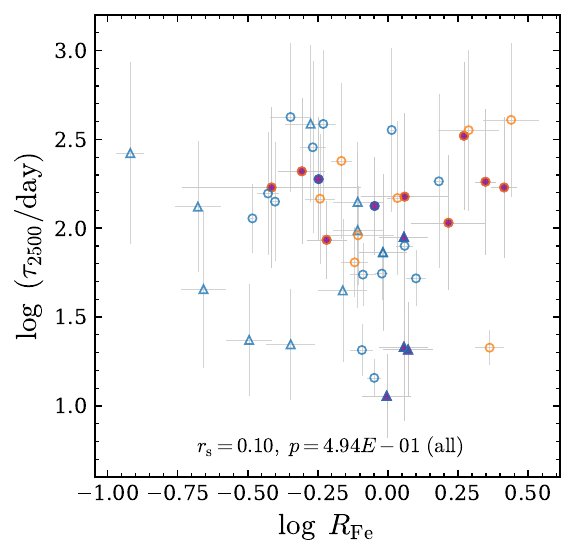}
\caption{Correlations between \tauuv\ and the observational AGN properties, \lv\ and \rfe\ from the left to right panels, respectively. Circles and triangles denote AGNs from \cite{Shen_2024} and \cite{Burke_2021}, respectively, for our sample of 115 AGNs. Symbols are color coded, with blue, orange, and gray representing \hb, \ion{Mg}{2}, and \ion{C}{4}-based virial \mbh, respectively. \rs\ and its \p\ are listed at the bottom of each panel. The linear regression is performed for significant correlations with $\p<0.01$. The black solid line and the gray shaded region represent the best-fit linear relation and its $3\sigma$ uncertainty, respectively. The slope of the best-fit line ($k$) and $\chi_\nu^2$ are labeled in the legend. The purple color in the right panel represents 14 NLS1s. Error bars are shown in lighter lines.}
\label{fig:5}
\end{figure*}

\begin{figure*}
    \centering
    \includegraphics[width=0.32\textwidth]{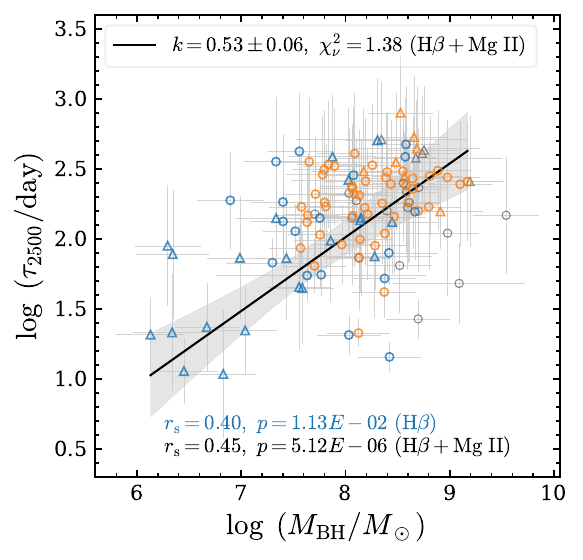}
    \hfill
    \includegraphics[width=0.32\textwidth]{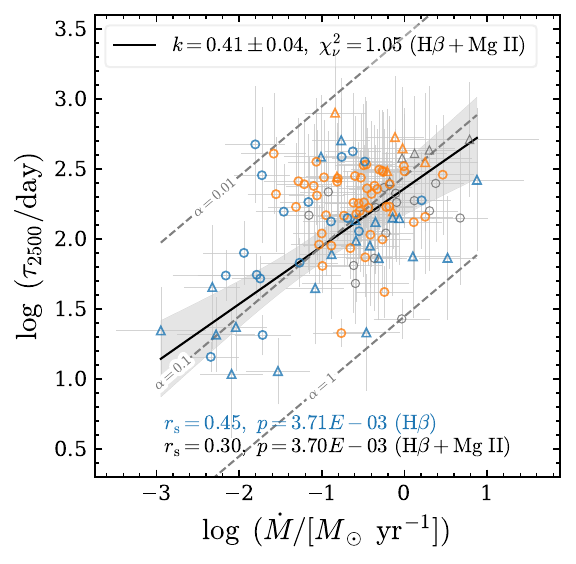}
    \hfill
    \includegraphics[width=0.32\textwidth]{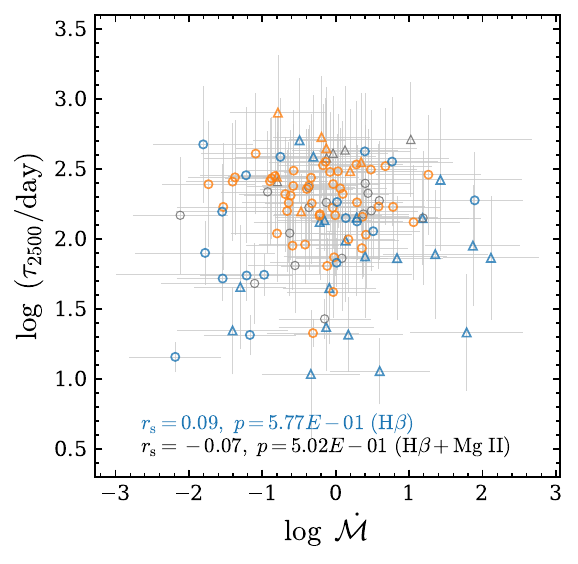}
\caption{Correlations between the damping timescale at rest-frame 2500~\AA\ and the SMBH accretion properties for the selected sample of 94 AGNs with \hb\ or \ion{Mg}{2}-based virial \mbh. Symbols are the same as in Figure~\ref{fig:5}. In the middle panel, the thermal timescale of the standard thin disk for different values of $\alpha$ ($0.01,~0.1,~1$) is represented by gray dashed lines, where $\alpha = 1$ corresponds to the dynamical timescale.}
\label{fig:6}
\end{figure*}

\begin{figure*}
\plotone{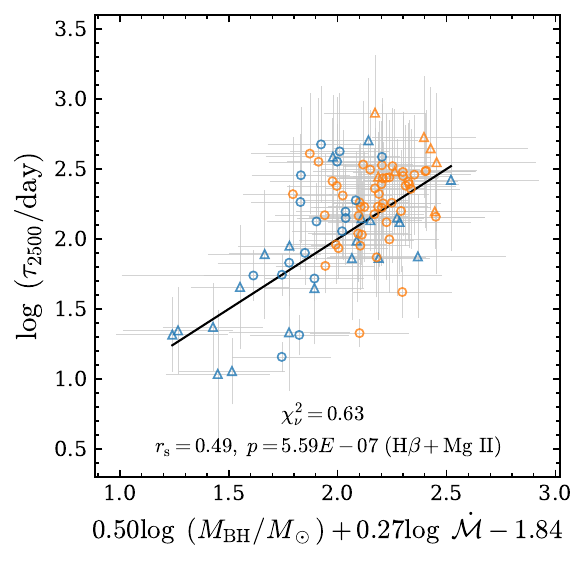}
\caption{Multivariate linear regression of $\log\ (\tauuv/\rm day)$ as a function of SMBH accretion quantities, $\log\ (\mbh/M_\odot)$ and $\log\ \mdotdim$  for our selected sample of 94 AGNs with \hb\ or \ion{Mg}{2}-based virial \mbh. Symbols are the same as in Figure~\ref{fig:5}.}
\label{fig:7}
\end{figure*}

For our sample of 115 AGNs, Figure~\ref{fig:5} shows the relations between the damping timescale at rest-frame 2500~\AA\ and various observational AGN properties, \lv\ and \rfe\ from the left-to-right panels, respectively. \lv\ ranges from $\sim 10^{41}$ to $\rm 10^{46}\ erg~s^{-1}$. log\rfe\ ranges from $\sim -1.0$ to $0.5$. The timescale \tauuv\ shows significant positive correlations with \lv\ (see left panel in Figure~\ref{fig:5}), with $\rs=0.43$, $\p=1.98\times10^{-6}$. The Spearman correlation test shows no significant correlation between \tauuv\ and \rfe, with $\rs=0.10$ and $\p=0.49$ (see right panel in Figure~\ref{fig:5}). For 14 NLS1s, the Spearman correlation test also shows no significant correlation between them ($\rs=0.04$ and $\p=0.89$). 

The \texttt{lmfit}\footnote{\url{https://github.com/lmfit/lmfit-py}} \citep{Newville_2025} is used to perform a linear regression, in the form of $y=kx+b$, for significant correlations with $\p<0.01$ (shown in bold in Table~\ref{tab:1}). The Markov Chain Monte Carlo (MCMC) sampler \texttt{emcee} in the \texttt{lmfit} is used to obtain the posterior distribution of the linear regression parameters with a uniform prior and $\texttt{nsample}=2000$ and $\texttt{nburn}=500$. The log-likelihood is $-0.5\sum_i^N[\frac{[y_i-(kx_i+b)]^2}{\sigma_i^2}+\ln(2\pi\sigma_i^2)]$, where $\sigma_i^2=\sigma_{y_i}^2+(k\sigma_{xi})^2$, $x_i$ and $y_i$ are independent and dependent variables, $\sigma_{x_i}$ and $\sigma_{y_i}$ are their corresponding uncertainties, and $k$, $b$, $N$ are the slope, intercept, and number of data points, respectively. The best-fit values are taken as the posterior medians, and the uncertainties are half the difference between the 16th and 84th percentiles of the posterior distribution. 

Since most AGNs do not have reported uncertainties for \lv, we add 0.1 dex in quadrature to \lv\ \citep[corresponding to a fractional uncertainty of $\sim$ 20\%,][]{Burke_2021}. Using \texttt{lmfit}, the best-fit relation is $\tauuv \propto \lv^{0.32\pm0.04}$ for our sample of 115 AGNs (see the left panel in Figure~\ref{fig:5}). Considering errors in all variables, the reduced chi-square defined as $\chi_\nu^2=\frac{1}{N-2}\sum_i^N\frac{[y_i-(kx_i+b)]^2}{\sigma_i^2}$ is also shown in the left panel in Figure~\ref{fig:5}. The large $\chi_\nu^2$ values suggest that there must be additional parameters contributing to the observed scatter. The slope of $0.32\pm 0.04$ for the relation between \tauuv\ and \lv\ is consistent with $0.29\pm 0.08$ by \cite{Kelly_2009}, larger than $-0.05\pm0.17$ by \cite{Kozlowski_2016}, and is smaller than $0.46\pm 0.09$ by \cite{Lu_2019}. A comprehensive comparison of literature results on the dependence of AGN optical variability on the luminosity, \mbh, and \mdotdim\ is given in Table 1 of \cite{Paolillo_2025}.

For our selected sample of 94 AGNs with \hb\ or \mgii-based \mbh, Figure~\ref{fig:6} shows the correlations between the damping timescale in the rest-frame 2500~\AA\ and the SMBH accretion properties, \mbh, \mdot, and \mdotdim, from left to right panel, respectively. \mbh\ spans a range of $\sim 10^6 -10^9\msun$. $\log~(\mdot/[M_\odot~\rm yr^{-1}])$ spans a range of $\sim -3.0$ to $1.0$. log~\mdotdim\ spans a range of $\sim -3.0$ to $2.0$. The timescale \tauuv\ exhibits significant positive correlations with \mbh, with $\rs=0.45$ and $\p=5.12\times10^{-6}$. The relation of $\tauuv$ with $\lv$ ($\rs=0.48$ and $\p=7.55\times10^{-7}$) is slightly stronger than that with \mbh. The timescale \tauuv\ shows a significant positive correlation with \mdot, with $\rs=0.30$ and $\p=0.004$. No significant correlation is found with \mdotdim, with $\rs=-0.07$ and $\p=0.50$. Considering errors on both axes, \texttt{lmfit} yields the best-fit relations as $\tauuv \propto \mbh^{0.53\pm0.06}$ and $\tauuv \propto \mdot^{0.41\pm0.04}$, respectively (see the left and middle panels in Figure~\ref{fig:6}). The slope of $0.53\pm 0.06$ for the correlation between \tauuv\ and \mbh\ is larger than the value of $0.21\pm 0.07$ found by \cite{Macleod_2010} and $0.38\pm0.05$ by \cite{Burke_2021}, and consistent with $0.56 \pm 0.14$ found by \cite{Kelly_2009}, $0.60\pm 0.09$ by \cite{Lu_2019}, and $0.6\pm 0.1$ by \cite{Petrecca_2026}\footnote{Note that the break timescale measured in \cite{Petrecca_2026} is from a bending power-law and thus it is not the DRW timescale.}. As discussed above, such discrepancies likely stem from differences in the adopted rest-frame wavelength, sample selection, and light-curve modeling methods \cite[see also Table~1 of ][]{Paolillo_2025}.

It is known that there are correlations among \lv, $1+z$, \mbh, and \mdotdim. For our selected sample of 94 AGNs, \lv\ shows strong correlations with both \opz\ ($\rs=0.75, \p=7.01\times10^{-18}$), and \mbh\ ($\rs=0.72, \p=3.11\times10^{-16}$). To identify the dominant correlations between DRW parameters and AGN properties, we therefore employ partial correlation analysis \footnote{\url{https://pingouin-stats.org/build/html/generated/pingouin.partial_corr.html}}. The partial correlation coefficient ($r_{xy|z}$) between variables $x$ and $y$, while controlling for variable $z$ is derived from the Spearman correlation coefficients between the three pairs ($r_{xy}, r_{xz}, r_{yz}$) and is computed as
\begin{equation}
 r_{xy|z} = \frac{r_{xy} - r_{xz} r_{yz}}{\sqrt{\left(1 - r_{xz}^2\right) \left(1 - r_{yz}^2\right)}}.
\end{equation}

The partial correlation between \tauuv\ and \lv, after controlling for $1+z$, is significant ($\rs=0.27, \p=0.003$). In contrast, the partial correlation between \tauuv\ and $1+z$, after controlling for \lv, is not significant ($\rs=0.01, \p=0.93$). The partial correlation between \tauuv\ and \lv\ or $1+z$ becomes not significant after controlling for \mbh\ ($p>0.01$), implying that these apparent correlations are induced by the dependence of $L_{5100}$ on $M_{\rm BH}$ and selection effects in luminosity and mass as a function of redshift. This result is consistent with the multivariate regression analysis of \cite{Macleod_2010}. Regarding the accretion properties, the partial correlation between \tauuv\ and \mbh, after controlling for \mdot, remains significant ($\rs=0.43,\p=1.40\times10^{-5}$). The partial correlation between \tauuv\ and \mdot, after controlling for \mbh, is less significant ($\rs=0.27, \p=0.01$). These results suggest that the relation of \tauuv\ with \mbh\ is more  fundamental than the relation between \tauuv\ and \mdot. 

Since \mbh\ and \mdotdim\ are two key parameters describing the SMBH accretion process, we employ multivariate regression analysis to investigate the relation of \tauuv\ with \mbh\ and \mdotdim\ or \rfe. Accounting for the uncertainties in \tauuv, \mbh, and \mdotdim, and fitting with \texttt{lmfit}, we obtain $\tauuv \propto \mbh^{0.50\pm0.07}\mdotdim^{0.27\pm0.04}$ (see Figure~\ref{fig:7}). This indicates that \tauuv\ depends on both \mbh\ and \mdotdim. To assess whether including \mdotdim\ significantly improves the fit, we perform an F-test and find that the probability of no improvement  is $1.11 \times 10^{-16}$. The improvement therefore exceeds $3\sigma$ significance ($1-P > 99.73\%$) \citep[e.g.,][]{Chen_2022}. The derived slope of $0.50\pm 0.07$ is consistent with $0.53\pm 0.06$ from the bivariate \tauuv\ -- \mbh\ relation. The correlation results between \tauuv\ and the AGN properties are summarized in Table~\ref{tab:1}, and the partial correlation analyses of the variability timescale are presented in Table~\ref{tab:2}.

\subsection{The Relations between the Variability Amplitude and the AGN Properties}

\begin{deluxetable*}{lcccc}
\setlength{\tabcolsep}{3.5pt}
\tablecaption{Correlations between \sigmauv\ and various AGNs properties.
\label{tab:3}
}
\tablehead{
\colhead{Properties} &
\colhead{$\rs$} &
\colhead{$\p$} &
\colhead{$k$} &
\colhead{$\chi_\nu^2$}
}
\startdata
$\log \lv$ & \boldmath{$-0.24$} & \boldmath{$9.61\times10^{-3}$} & $0.02\pm0.02$ & 2.04 \\
$\log \rfe$ & 0.10 & 0.50 & --- & --- \\
$\log \mbh$ & 0.06 & 0.57 & --- & --- \\
$\log \mdot$ & $-0.16$ & 0.12 & --- & --- \\
$\log \mdotdim$ & $-0.19$ & 0.06 & --- & --- \\
$(\log \mbh,\ \log \mdotdim)$ & 0.16 & 0.12 & $(-0.09\pm0.04,\ -0.13\pm0.02)$ & 1.15 \\
$(\log \mbh,\ \log \rfe)$ & 0.23 & $0.13$ & $(0.21\pm0.03,\ 0.18\pm0.09)$ & 1.86 \\
$(\log \mbh,\ \log \lv)$ & 0.18 & 0.08 & $(0.24\pm0.03,\ -0.12\pm0.03)$ & 1.26 \\
\enddata
\tablecomments{
Listed are the Spearman correlation coefficient (\rs), $p$-value (\p), best-fit slope ($k$), and reduced chi-square ($\chi_\nu^2$) for the correlations between $\log \sigmauv$ and various AGN properties. Significant correlations with $\p < 0.01$ are highlighted in bold.
}
\end{deluxetable*}

\begin{deluxetable*}{lccc}
\setlength{\tabcolsep}{3.5pt}
\tablecaption{Summary of the partial correlation analyses for \sigmauv. \label{tab:4}}
\tablehead{
\colhead{Parameter Pair} & \colhead{Controlled Variable} & \colhead{$\rs$} & \colhead{$\p$} 
}
\startdata
$\sigmauv$ -- $\mdotdim$ & $\mbh$ & $-0.20$ & 0.06 \\
$\sigmauv$ -- $\mbh$ & $\mdotdim$ & $-0.07$ & 0.52 \\
$\sigmauv$ -- $\lv$ & $\mdotdim$ & $-0.04$ & 0.72 \\
$\sigmauv$ -- $1+z$ & $\mdotdim$ & $0.07$ & 0.52 \\
$\sigmauv$ -- $\mdot$ & $\mdotdim$ & $-0.04$ & 0.69 \\
$\sigmauv$ -- $\lv$ & $1+z$ & $-0.16$ & 0.09 \\
$\sigmauv$ -- $1+z$ & $\lv$ & 0.01 & 0.91 \\
\enddata
\tablecomments{Listed are the Spearman correlation coefficient (\rs) and $p$-value (\p) of the partial correlation analyses.}
\end{deluxetable*}

\begin{figure*}
    \centering
    \includegraphics[width=0.32\textwidth]{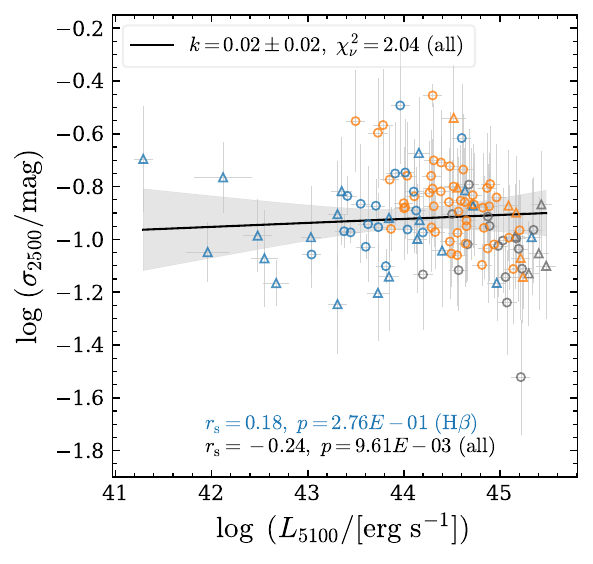}
    \includegraphics[width=0.32\textwidth]{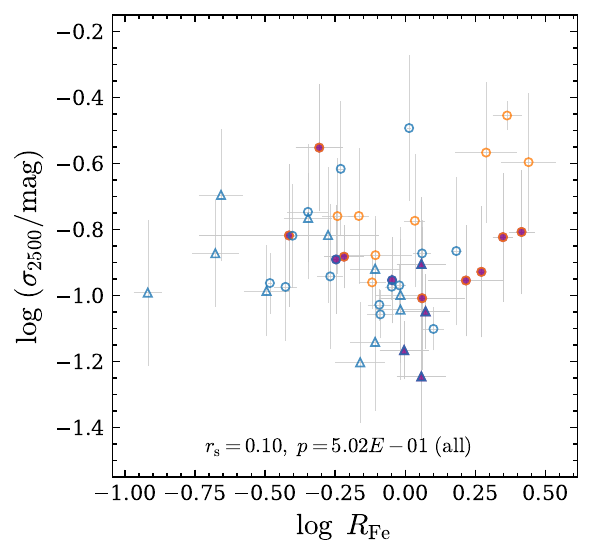}
\caption{Correlations between \sigmauv\ and the observational AGNs properties for our sample of 115 AGNs. Symbols are the same as in Figure~\ref{fig:5}.}
\label{fig:8}
\end{figure*}

\begin{figure*}
    \centering
    \includegraphics[width=0.32\textwidth]{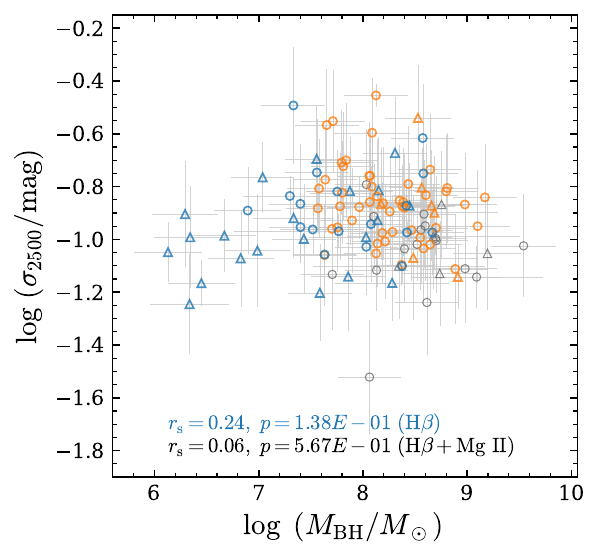}
    \hfill
    \includegraphics[width=0.32\textwidth]{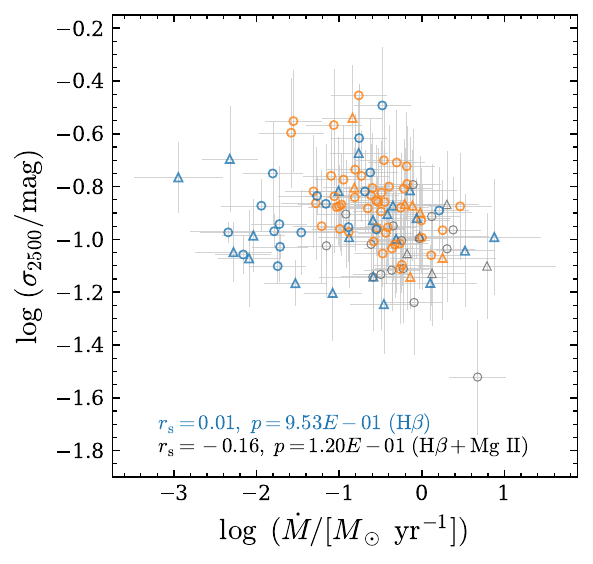}
    \hfill
    \includegraphics[width=0.32\textwidth]{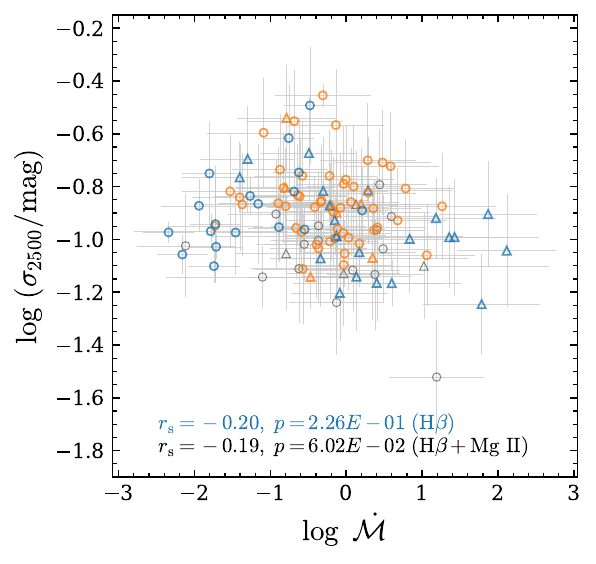}
\caption{Correlations between \sigmauv\ and the SMBH accretion properties for our sample of 94 AGNs with \hb\ or \ion{Mg}{2}-based virial \mbh. Symbols are the same as in Figure~\ref{fig:5}.}
\label{fig:9}
\end{figure*}

For our sample of 115 AGNs, Figure~\ref{fig:8} shows the correlations between \sigmauv\ and various observational AGN properties, \lv\ and \rfe\ from left to right, respectively. \sigmauv\ and \lv\ exhibit no definitive evidence of a correlation, with $\rs=-0.24$ and $\p=0.0096$ ($\rs=-0.36$ and $\p=9.22\times10^{-4}$ for the SDSS-RM subsample). The lack of a definitive correlation between \sigmad\ and \lv\ is also supported by the flat fitted relation, $\sigmauv \propto L_{5100}^{0.02\pm0.02}$. The smaller \taud\ values observed in low-luminosity sources, together with the correlation between \taud\ and \sigmad\ (Figure~\ref{fig:3}), may contribute to the lack of a definitive correlation between \sigmad\ and \lv. A Spearman correlation test indicates that there is no significant correlation between \sigmauv\ and $1+z$ for our sample of 115 AGNs ($\rs=-0.18, \p=0.05$). While for the SDSS-RM subsample, the significance appears larger ($\rs=-0.38$, $\p=4.24\times10^{-4}$), a partial correlation between $\sigmauv$ and $\lv$ after controlling for $z$ yields $\rs=-0.08$, $\p=0.49$, and between $\sigmauv$ and $z$ after controlling for $\lv$ yields $\rs=-0.15$, $\p=0.17$. Thus, these correlations appear again as a byproducts of the dependence of luminosity on redshift. The correlation between $\sigmauv$, $\lv$ and $z$ cannot be clearly disentangled even for the SDSS-RM subsample. For \sigmauv\ shows no significant correlation with \rfe\ ($\rs=0.10, \p=0.50$). 

For our selected sample of 94 AGNs with \hb-\ or \mgii-based \mbh, Figure~\ref{fig:9} shows the correlations between \sigmauv\ and the SMBH accretion properties, \mbh, \mdot, and \mdotdim\ from left to right, respectively. \sigmauv\ shows no significant correlation with \mbh, with $\rs=0.06$ and $\p=0.57$, which is consistent with the result of \cite{Petrecca_2026}. The correlation between variability amplitude and \mbh\ has been debated in previous studies (e.g., \citealp{Arevalo_2023}; also see the review of \citealp{Paolillo_2025}). A positive correlation has been reported in several works \citep[e.g.,][]{Macleod_2010, Lu_2019, Kovacevic_2025}. \cite{Arevalo_2023} found that the correlation between variability amplitude and \mbh\ depends on the timescale of the variability studied. They found a negative correlation between variability amplitude and \mbh\ and found that this anticorrelation is stronger for shorter-timescale variability. \cite{Benati_Goncalves_2025} found that the correlation between variability amplitude and \mbh\ depends on the redshift $z$: it is positive at low $z$, insignificant at intermediate $z$, and becomes an anticorrelation at high $z$. 

\sigmauv\ also shows a marginally significant trend with \mdotdim\ ($\rs=-0.19, \p=0.06$). It is consistent with previous others studies on the anticorrelation between $\Delta m$ and the Eddington ratio \citep[e.g.,][]{Wilhite_2008, Macleod_2010,Ai_2010, Zuo_2012}. AGN with higher Eddington ratios may possess smoother fueling material, or become more radiatively efficient with smoother energy output, and suppress stochastic variability, which would lead to this anticorrelation \citep[e.g.,][]{Wu_2024, Benati_Goncalves_2025}. 

Regarding the observational properties, the partial correlation between \sigmauv\ and \lv, after controlling for $1+z$, is $\rs=-0.16$ with $\p=0.09$. The partial correlation between \sigmauv\ and $1+z$, after controlling for \lv, is $\rs=0.01$ with $\p=0.91$. Regarding the accretion properties, the partial correlation between \sigmauv\ and \mdotdim, after controlling for \mbh, is $\rs=-0.20$ with $\p=0.06$. The partial correlation between \sigmauv\ and \mbh, after controlling for \mdotdim, is $\rs=-0.07$ with $\p=0.52$. These results suggest that \sigmauv\ is more strongly correlated with \mdotdim\ than with \mbh. The partial correlation between \sigmauv\ and \lv\ becomes insignificant ($\p=0.72$) after controlling for \mdotdim, implying that the apparent correlations between \sigmauv\ and \lv\ are likely driven by their mutual dependence on \mdotdim. 

The multivariate regression analysis is also used to investigate the relation of \sigmauv\ with \mbh\ and \mdotdim\ or \rfe. Considering the uncertainties of \sigmauv, \mbh\ and \mdotdim, using \texttt{lmfit}, the multivariate regression yields $\sigmauv \propto \mbh^{-0.09\pm0.04}\mdotdim^{-0.13\pm0.02}$ ($p=0.12, \chi_\nu^2=1.15 $). This suggests that the normalized variability amplitude has dependence on the accretion rate traced by \mdotdim, as well as \mbh. With fixed \mdotdim, there is a negative relation between \sigmauv\ and \mbh\ with a slope of $-0.09\pm0.04$. This is not consistent with the result by \cite{Li_2008}. However, because both \mbh\ and \mdotdim\ are related to the continuum luminosity, the apparent dependencies of \sigmauv\ on these quantities may be induced, at least in part, by their mutual dependence on the continuum luminosity. The multivariate regression yields $\sigmauv \propto \mbh^{0.24\pm0.03}\lv^{-0.12\pm0.03}$ ($p=0.08, \chi_\nu^2=1.26 $). \sigmauv\ has a negative relation with \lv\ and positive relation with \mbh, which is consistent with the result by \cite{Li_2008} (see their Figure~3). Using \rfe\ instead of \mdotdim, the multivariate regression yields $\sigmauv \propto \mbh^{0.21\pm0.03} \rfe^{0.18\pm0.09}$ ($p=0.13, \chi_\nu^2=1.86 $). These results are tentative trends rather than significant correlations; larger samples are required to confirm these possible trends. The results of the correlations between \sigmauv\ and the AGN properties are presented in Table~\ref{tab:3}. The partial correlation analyses of the variability amplitude are summarized in Table~\ref{tab:4}.

\section{Discussion} \label{sec:4} 

\subsection{The Relation between the Damping Timescale and the Thermal Timescale} 

Since the optical/UV continuum variability of AGNs originates from the SMBH accretion processes, the damping timescale is expected to be linked to the characteristic timescales of the accretion disk \citep[e.g.,][]{Collier_2001, Kelly_2009, Ren_2024, Zhou_2024}. However, the physical origin of AGN variability, particularly on different timescales, remains under debate. We investigate the connections between the damping timescale at 2500~\AA\ (\tauuv) from the long-term light curves and the dynamical timescales as well as the thermal timescale of the standard thin-disk model. It is worth noting that the standard thin-disk model applies to the regime of $0.01<\lambda_{\rm Edd}<0.3$ \citep[e.g.,][]{Son_2025}. In our sample, three objects have $\lambda_{\rm Edd}<0.01$, and 18 objects have $\lambda_{\rm Edd}>0.3$, meaning that 94 out of 115 (82\%) objects lie within the regime where the standard thin-disk model is expected to apply. Assuming a Keplerian velocity at a distance of $R$ from the center of the accretion disk, the dynamical timescale and the thermal timescale can be calculated as follows \citep[e.g.,][]{Netzer_2025}:
\begin{equation} 
\label{eq:t_dyn}
t_{\rm dyn} = 0.005 M_8 \left( \frac{R}{R_{\rm g}} \right)^{3/2} \rm ~day,
\end{equation}
\begin{equation}
t_{\rm th} = \frac{t_{\rm dyn}}{\alpha},
\label{eq:t_th}
\end{equation}
where $M_8=\mbh/(10^8M_\odot)$, $R_{\rm g} = G\mbh/c^2$ is the gravitational radius, and $\alpha$ is the viscosity parameter. The relation between $t_{\rm th}$ and the viscous timescale $t_{\rm vis}$ is $t_{\rm th}\sim (H/R)^2 t_{\rm vis}$, where $H$ is the disk thickness \citep[e.g.,][]{Collier_2001, Gaskell_2006}. For the distance $R$ from the center of the accretion disk, where emissions peak at a wavelength $\lambda$ , the disk temperature profile must be specified. The temperature of the standard thin disk is calculated as $T(R)= \left( \frac{3G \mbh \mdot f(R)}{8 \pi \sigma R^3 } \right) ^{1/4}$ \citep{Frank_2002, Starling_2004, Cackett_2007, Netzer_2013}, where $\sigma$ is the Stefan–Boltzmann constant, and $f(R)=1-(R_{\rm in}/R)^{1/2}$, $R_{\rm in}$ denotes the inner radius of the accretion disk, and $f(R)=1$ is assumed. Thus, the temperature, being $T \propto R^{-3/4}$, can be expressed as \citep[e.g.,][]{Netzer_2013}
\begin{equation}
T(R) \approx 8.6 \times 10^5 M_8^{-1/2} \dot{M}_1^{1/4} \left( \frac{R}{R_{\rm g}} \right)^{-3/4} \rm ~K ,
\end{equation}
where the mass accretion rate $\dot{M}_1=\mdot/(M_\odot \rm~yr^{-1})$. According to Wien's displacement law, $\lamrest T=0.2898~\rm cm ~K$ \citep[e.g.,][]{Smith_2018, Son_2025}, the rest-frame 2500~\AA\ corresponds to a temperature of $11,592~\rm K$. The normalized distance can be calculated as follows:
\begin{equation}
\frac{R}{R_{\rm g}} \approx 312M_8^{-2/3}\dot{M}_1^{1/3} \left(\frac{\lamrest}{2500~\rm \AA}\right)^{4/3}.
\end{equation}
Therefore, the thermal timescale under the standard thin-disk model can be expressed as follows:
\begin{equation} 
\label{eq:t_th1}
t_{\rm th} \approx 28\alpha^{-1} \dot{M}_1^{1/2} \left(\frac{\lamrest}{2500~\rm \AA}\right)^2 \rm ~day, \nonumber
\end{equation}

In the middle panel in Figure~\ref{fig:6}, the theoretical thermal timescale for different values of $\alpha$ ($0.01,~0.1,~1$) is represented by gray dashed lines, where $\alpha=1$ corresponds to the dynamical timescale. Data points lie primarily between the thermal timescale with $\alpha=0.01$ and $\alpha=1$. This implies that the damping timescale \tauuv\ may not be related to the dynamical timescale because almost all of AGNs show larger \tauuv\ than the dynamical timescale. \tauuv\ is likely related to the thermal timescale. The relation of $\tauuv \propto \mdot^{0.41\pm0.04}$ is consistent with the thermal timescale expected from the standard thin-disk model within $3\sigma$ uncertainties, but with substantial scatter. The scatter in this relation may arise from the variations in the viscosity parameter $\alpha$ or a different temperature profile. It implies that a different viscosity parameter $\alpha$ or a different temperature profile from the standard thin-disk model \citep{Li_2008} should be adopted when analyzing the long-term continuum variability timescale.

The rest-frame wavelength dependence of \taud\ we adopted, $\taud\ \propto \lamrest^{0.17}$ \citep{Macleod_2010}, is inconsistent with the thermal timescale of the standard thin-disk model ($\taud\ \propto \lamrest^{2}$; Equation \ref{eq:t_th1}). To test the robustness of our results, we also considered the steeper relation $\taud\ \propto \lamrest^{1.42}$ from \cite{Son_2025}; the main results remain unchanged. This insensitivity suggests that our conclusions do not depend on the specific choice of the $\taud-\lamrest$ relation. The discrepancy with the standard model may partly arise because broadband photometry integrates over a range of disk radii, which can attenuate the apparent dependence of \taud\ on AGN properties, and because emission-line contamination can bias the measured lags. Spectroscopic studies also find similar inconsistencies, which may indicate deviations of the temperature profile from the standard thin-disk model \citep[e.g.,][]{Son_2025, Kim_2026}. For instance, assuming a single universal structure function, \cite{Son_2025} found that the characteristic timescale scales as $L_{\rm bol}^{0.50\pm0.03} \lamrest^{1.42\pm0.09}$ for SDSS-RM quasars; the slope of $1.42 \pm 0.09$ deviates from the value of 2 expected in the standard model, possibly reflecting differences in accretion efficiency. On the other hand, photometric continuum RM studies generally find a temperature profile $T \propto R^{-3/4}$, but with a normalization $\sim 3$ times larger than theoretical predictions \citep[e.g.,][and references therein]{Guo_2022}. This apparent excess may not be intrinsic; it could result from obscuration effects adding extra variability to the continuum light curves \citep{Lewin_2025} or from color corrections due to scattering in the disk atmosphere \citep[e.g.,][]{Hall_2018}. These complexities affect continuum lags and the temperature profile, but not our main conclusions on the DRW \taud\ versus AGN properties.

\subsection{The Relation between the Disk-emitting Radius at 2500~\AA\ and the Black Hole Mass} 

\begin{figure*}
\epsscale{0.8}
\plotone{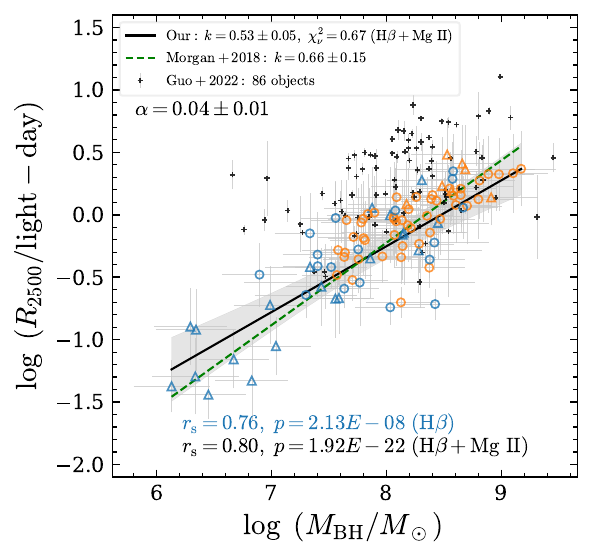}
\caption{Correlation between the disk-emitting radius at rest-frame 2500~\AA\ and the SMBH mass for our selected sample of 94 AGNs with \hb\ or \mgii-based \mbh. Circles and triangles represent AGNs from \cite{Shen_2024} and \cite{Burke_2021}, respectively. The symbols are color coded, with blue and orange representing \hb- and \ion{Mg}{2}-based virial \mbh, respectively. The black plus symbols represent the continuum RM result of AGNs from \cite{Guo_2022}. The green dashed line is the microlensing-based \ruv-\mbh\ relation of \cite{Morgan_2018}. The best-fit value is $\alpha=0.04\pm0.01$. The black solid line is the best-fit relation using the value of $\alpha=0.04$, the best-fit slope is $k=0.53\pm0.05$. The gray shaded region is the $3\sigma$ uncertainty.}
\label{fig:10}
\end{figure*}

Using various methods, a relation between the disk-emitting radius at rest-frame 2500~\AA\ (\ruv) and the black hole mass is reported by some studies \citep[e.g.,][]{Morgan_2018, Burke_2021, Guo_2022}.
From the microlensing observations, sources smaller than the Einstein radius of the stars have larger variability amplitudes than larger sources, \cite{Morgan_2018} found a scaling relation of $\ruv \propto \mbh^{0.66\pm0.15}$, where the virial SMBH \mbh\ were estimated from different broad lines. 
Using the DRW-based approach and assuming that the damping timescale corresponds to the thermal timescale, \cite{Burke_2021} found that the variability data are consistent with the relation of \cite{Morgan_2018}. Based on the continuum RM method and the standard thin-disk temperature profile of $T \propto R^{-3/4}$, \cite{Guo_2022} calculated the disk size at 2500~\AA\ and reported a similar correlation, but with different slope, $\ruv \propto \mbh^{0.72\pm0.13}$. 

For our selected sample of 94 AGNs with \hb\ or \mgii-based \mbh, assuming that the damping timescale corresponds to the thermal timescale, combining Equation~(\ref{eq:t_dyn}) and Equation~(\ref{eq:t_th}) yields 
\begin{equation}
\ruv \approx 4.2\times10^{-4}\alpha^{2/3} \left( \frac{\tauuv}{\rm day} \right)^{2/3} \left( \frac{\mbh}{M_\odot} \right)^{1/3}~\rm lt-day.
\label{eq:13}
\end{equation}
The uncertainty of $\ruv$ is calculated by propagating the uncertainties of $\tauuv$ and $\mbh$.

The choice of mean value of $\alpha$ affects the intercept of the $\log \ruv-\log \mbh$ relation, but does not influence its slope. We leave $\alpha$ as a free parameter and minimize $\chi_\nu^2$ between our calculated \ruv\ (using \tauuv, \mbh\ and Equation(\ref{eq:13})) and the predicted \ruv\ from the microlensing-based \ruv--\mbh\ relation of \cite{Morgan_2018}. Taking into account the uncertainties of \mbh\ and \tauuv, using \texttt{lmfit}, the best-fit viscosity parameter is $\alpha=0.04\pm0.01$. The uncertainty of $\alpha$ is underestimated, given the uncertainty in the \ruv--\mbh\ relation of \cite{Morgan_2018}. In thin and fully ionized disks, the observational evidence suggests a typical range of $\alpha \sim 0.1-0.4$ \citep{King_2007}, whereas the relevant numerical simulations tend to derive estimates for $\alpha$ which are an order of magnitude smaller. However, some studies found $\alpha \sim 0.01$ from quasar variability \citep[][and references therein]{Kelly_2009}.

For our selected sample of 94 AGNs with \hb\ or \mgii-based \mbh, adopting our fitting result of $\alpha=0.04$, we calculate \ruv, and show the \ruv--\mbh\ relation in Figure~\ref{fig:10}. \ruv\ ranges from $\sim 0.03$ to $3$ lt-day for \mbh\ ranging from $\sim 10^6$ to $10^9\msun$. There is a significant correlation between \ruv\ and \mbh\ with $r_s=0.80$ and $\p= 1.92\times 10^{-22}$. Using \texttt{lmfit}, we obtain $\ruv \propto \mbh^{0.53\pm0.05}$, which is consistent with the slope of $0.66\pm 0.15$ within uncertainty from the microlensing method by \cite{Morgan_2018} (green dashed line in Figure~\ref{fig:10}). In Figure~\ref{fig:10}, we also show the data of continuum RM from \cite{Guo_2022}. The continuum RM result appears to have a systematically larger \ruv\ compared to the microlensing-based result, suggesting that the standard thin-disk temperature profile may deviate from the canonical relation of the standard thin-disk model \cite[e.g.,][]{Li_2008}. Other possible explanations for this discrepancy include an improper treatment of relativistic effects and the need for colour-correction terms due to scattering.


\section{Conclusions} \label{sec:5}

\noindent Optical/UV continuum variability over baselines of up to 20 yr is investigated using the DRW model for 1214 AGNs compiled from \cite{Shen_2024} and \cite{Burke_2021}. For our sample of 115 AGNs with reliable DRW fitting and unbiased \taud, the DRW parameters, obtained from light curves, are converted to the rest-frame 2500~\AA\ using the wavelength dependence reported by \cite{Macleod_2010}. These parameters are used to probe relations with AGN properties. The main conclusions are summarized as follows:
\begin{itemize}
\item Through DRW modeling of long-term continuum light curves, we find that 671 out of 1214 (55\%) AGNs can be well fitted by the DRW model. We further select 115 from 671 (17\%) AGNs that satisfy the criteria for unbiased DRW parameters. A total of 556 AGNs are excluded from 671 AGNs, most of them showing large DRW timescales that are larger than one-fifth of the light-curve baseline. It underscores the importance of extending the light curves to a longer baseline in the future. Excluding 21 AGNs with \ion{C}{4}-based \mbh, the remaining 94 AGNs with \hb\ or \mgii-based \mbh, spanning a range of $\sim 10^6 -10^9\msun$, are used to investigate the variability relations with the SMBH accretion process. 

\item With bivariate correlation analysis, the rest-frame DRW timescale at 2500~\AA, \tauuv, shows significant positive correlations with the continuum luminosity at 5100~\AA, the SMBH mass, and the mass accretion rate, with $\rs=0.43$, $\rs=0.45$, and $\rs=0.30$, respectively. The best-fit relations are $\tauuv \propto \lv^{0.32\pm0.04}$, $\tauuv \propto \mbh^{0.53\pm0.06}$, and $\tauuv \propto \mdot^{0.41\pm0.04}$. The multivariate regression of \tauuv\ as a function of accretion-related quantities yields $\tauuv \propto \mbh^{0.50\pm0.07} \mdotdim^{0.27\pm0.04}$, indicating that the long-term DRW timescale depends on \mbh\ and the dimensionless accretion rate \mdotdim. The variability amplitude, \sigmauv, shows no definitive evidence of a correlation with the continuum luminosity at 5100~\AA\ for the full sample, although an anticorrelation is found for the SDSS-RM subsample. \sigmauv\ shows no significant correlations with \rfe, \mbh, and \mdotdim\ with $\p>0.01$. 

\item In the \tauuv-\mdot\ plane, data points lie primarily between the thermal timescale with $\alpha=0.01$ and $\alpha=1$. The best-fit relation of $\tauuv \propto \mdot^{0.41\pm0.04}$ is consistent with the thermal timescale expected from the standard thin-disk model within $3\sigma$ uncertainties, but with substantial scatter, implying that variations in the viscosity parameter $\alpha$ or temperature profile may contribute to the scatter in this relation.

\item Assuming that \tauuv\ corresponds to the thermal timescale, to be consistent with the microlensing-based \ruv-\mbh\ relation, the viscosity parameter is statistically estimated to be $\alpha=0.04\pm0.01$. Adopting $\alpha=0.04$, the disk-emitting radius at rest-frame 2500~\AA\ is calculated, ranging from about $0.03$ to $3$ lt-day for \mbh\ ranging from about $10^6$ to $10^9$ $\msun$. \ruv\ correlates with \mbh\ as $\ruv \propto \mbh^{0.53\pm0.05}$. 

\end{itemize}


\begin{acknowledgments}
We are grateful to the anonymous referee for the useful comments and suggestions, which helped us to improve the paper. This work has been supported by the National Science Foundation of China (No. 12573008). We also acknowledge data from \url{https://ariel.astro.illinois.edu/sdssrm/final_result/} and \url{https://github.com/burke86/taufit/tree/master/paper}.
\end{acknowledgments}

\begin{appendix}
\counterwithin{figure}{section}  
\renewcommand{\thefigure}{\Alph{section}.\arabic{figure}} 

\section{Four additional examples of the DRW fits with the smallest and largest timescale or amplitude} \label{sec:appendix}

\begin{figure*}
\epsscale{0.8}
\plotone{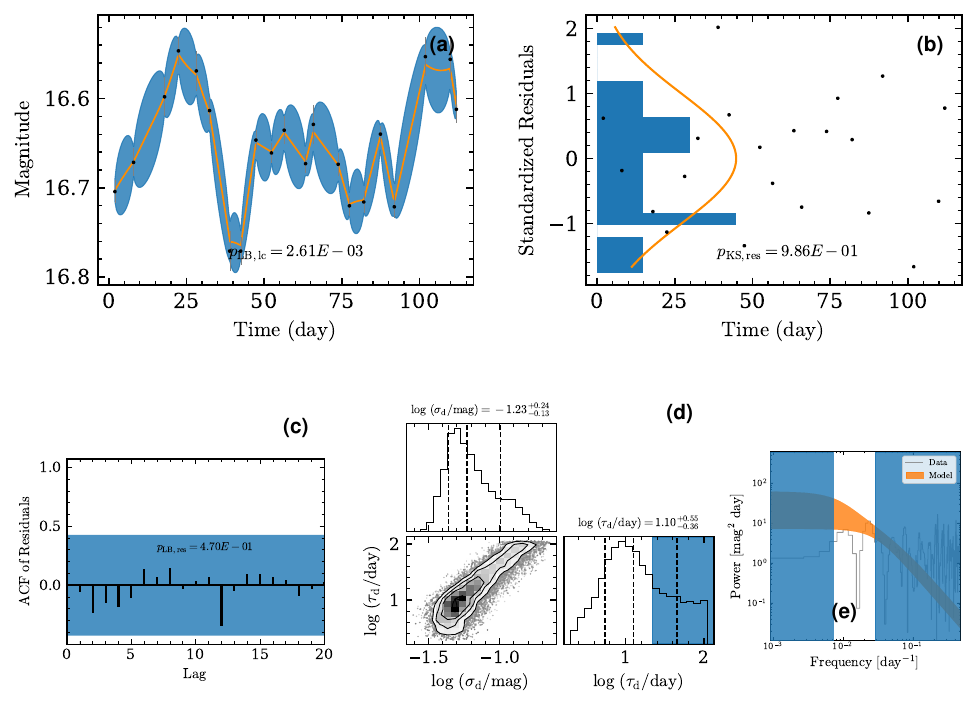}
\caption{DRW fit for the light curve of IC 4218 ($z=0.02$). IC 4218 has the smallest observed-frame \taud\ in our sample of 115 AGNs.}
\label{fig:11}
\end{figure*}

\begin{figure*}
\epsscale{0.8}
\plotone{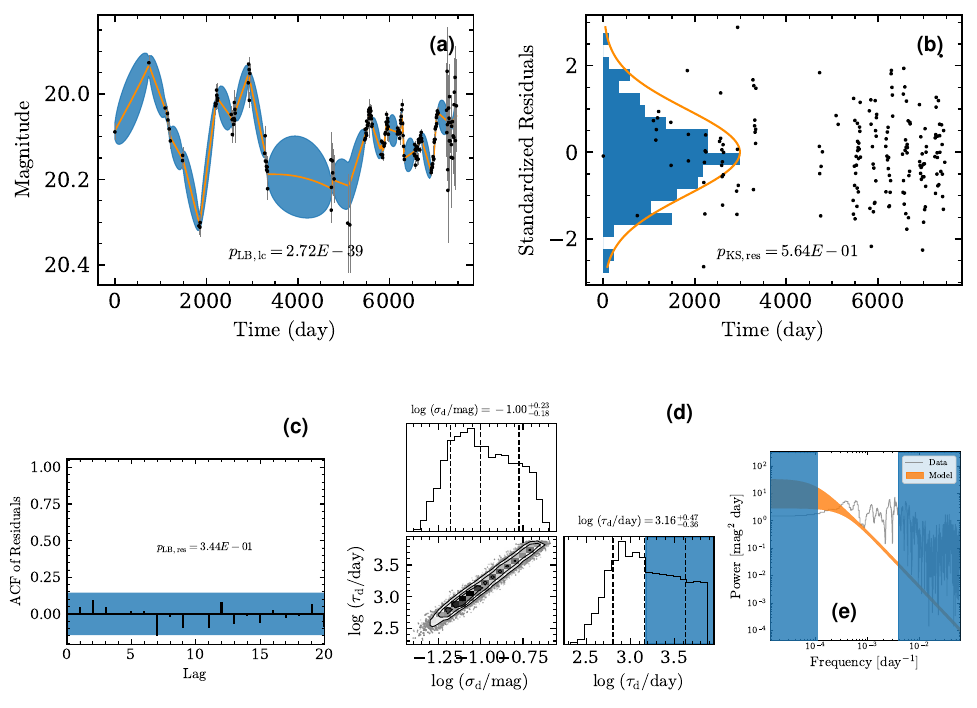}
\caption{DRW fit for the light curve of SDSS J025050.64+004503.2 ($z=2.05$). SDSS J025050.64+004503.2 has the largest observed-frame \taud\ in our sample of 115 AGNs.}
\label{fig:12}
\end{figure*}

\begin{figure*}
\epsscale{0.8}
\plotone{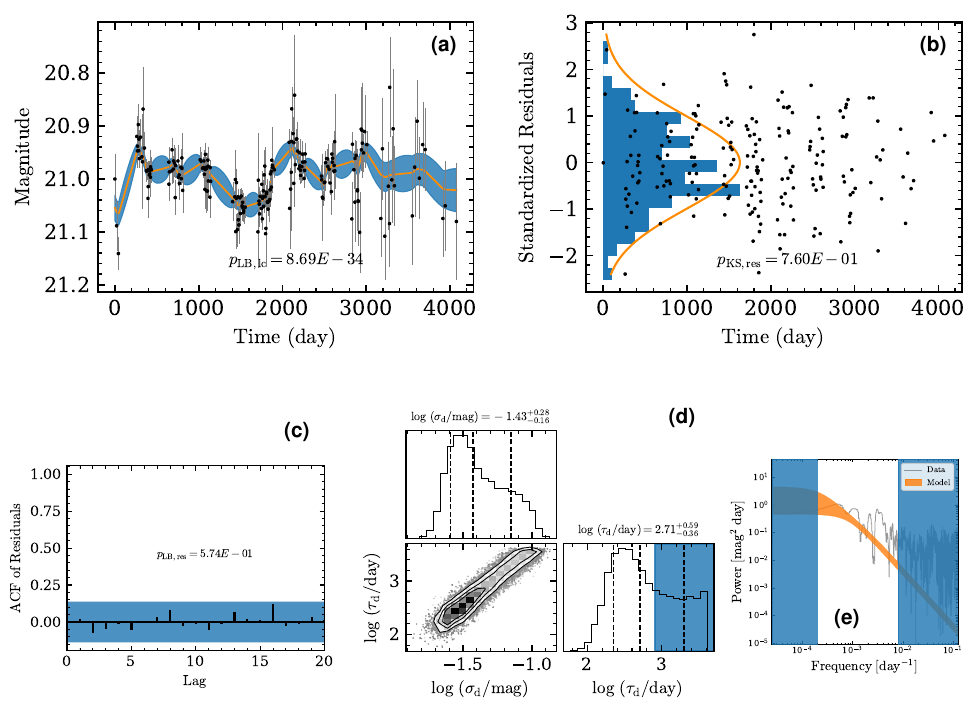}
\caption{DRW fit for the light curve of RM287 ($z=2.91$). RM287 has the smallest \sigmad\ in our sample of 115 AGNs.}
\label{fig:13}
\end{figure*}

\begin{figure*}
\epsscale{0.8}
\plotone{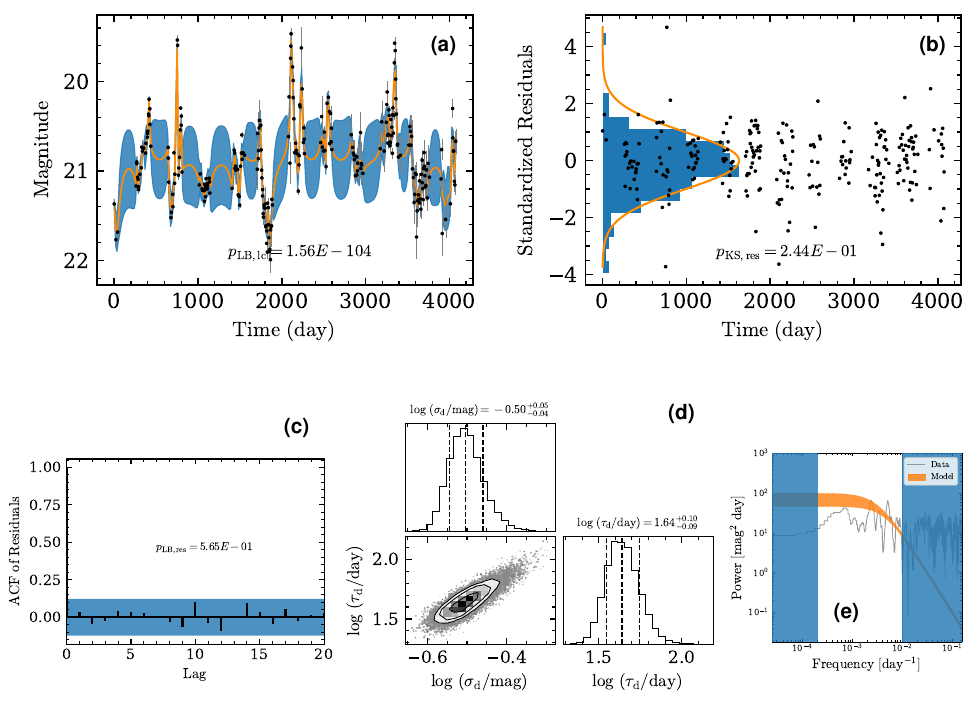}
\caption{DRW fit for the light curve of RM121 ($z=0.97$). RM121 has the largest \sigmad\ in our sample of 115 AGNs.}
\label{fig:14}
\end{figure*}

Figures~\ref{fig:11}-\ref{fig:14} show the DRW fit to the light curves of IC 4218 (with the smallest $\tau_{\rm d}$), SDSS J025050.64+004503.2 (with the largest $\tau_{\rm d}$), RM287 (with the smallest $\sigma_{\rm d}$), and RM121 (with the largest $\sigma_{\rm d}$), respectively.

\end{appendix}

\clearpage

\bibliography{reference}

\bibliographystyle{aasjournalv7}

\end{document}